# On statistical deficiency: Why the test statistic of the matching method is hopelessly underpowered and uniquely informative

Michael C. Nelson

RTI International

{michaelnelson} @ rti.org

**Abstract**

The random variate *m* is, in combinatorics, a basis for comparing permutations, as well as the solution to a centuries-old riddle involving the mishandling of hats. In statistics, m is the test statistic for a disused null hypothesis statistical test (NHST) of association, the matching method. In this paper, I show that the matching method has an absolute and relatively low limit on its statistical power. I do so first by reinterpreting Rae's theorem (1987), which describes the joint distributions of m with several rank correlation statistics under a true null. I then derive this property solely from *m*'s unconditional sampling distribution, on which basis I develop the concept of a deficient statistic—a statistic that is insufficient and inconsistent and inefficient with respect to its parameter. Finally, I demonstrate an application for *m* that makes use of its deficiency to qualify the sampling error in a jointly estimated sample correlation.

## 1 Introduction

Draw at random $n \geq 4$ bivariate observations $(X, Y)$ and convert to ranks, $(R[X], R[Y])$. Compute *m*, the number of paired observations with equal ranks:

$$m = \sum_{i}^{n} R[x_i] = R[y_i]$$

Then *m* is asymptotically distributed Poisson($\lambda = 1$), and the probability that the two populations are mutually independent (i.e., the *p*-value) is approximately Pois(1) = *m*. Reject the null hypothesis of independence when the *p*-value is greater than experimental $\alpha$; e.g., if the observed number of matches is $m = 4$, then $P(m > 3) \approx P(\text{Pois}(1) > 3) = .0417$, so reject at the conventional Type I error rate of $\alpha = .05$. This is one version of a null hypothesis significance test (NHST) of association called the matching method (cf. Vernon, 1936a, 1936b).

The virtues of the matching method (Evans, Hughes and Houston, 2002) are the ease of calculating its test statistic, the ease of determining *p*-values, and the ability to use the test with samples as small as $n = 4$ with no significant loss of power. The drawbacks are comparatively severe: the test provides no point estimate or uncertainty interval. Its *p*-values are discrete and therefore restrict the choice of empirical $\alpha$ to a few values. It has low statistical power and increasing sample size beyond a few observations does not meaningfully increase power. This last point follows directly from the asymptotic Poisson distribution of *m*, which ensures that *m* is a deficient statistic for the population correlation.

*Statistical deficiency*, as I define it here, is not simply the inadequacy or invalidity of a certain statistic for estimating a particular parameter. It is the quality of a statistic being highly insufficient *and* highly inconsistent *and* highly inefficient with respect to its parameter. I prove herein that the variable *m* displays these characteristics with respect to $\rho$ under a true null. The standard error of *m* will always be about the same as a Poisson-distributed variable with $\lambda = 1$, even as sample size goes to infinity. Consequently, there is an absolute limit on the amount of Fisher information *m* can convey about the population correlation, and a relatively low ceiling on the probability of rejecting a false null.

This is obviously no problem under a true null, when committing a Type II error is impossible. However, as I demonstrate, the standard error of *m* only *increases* as the population correlation departs from $\rho = 0$, and increasing standard errors mean decreasing information and decreasing power. For example, consider the (arbitrary) case where $n = 100$ and $\alpha = .05$. The conventional NHST in this case has 80 percent power to detect a population correlation of $\rho = .277$. The equivalent

NHST using the matching method has only 5 percent power to detect the same correlation.

The matching method was developed at the end of the nineteenth century (Vernon, 1936b) and largely fell out of use by social scientists by the mid-twentieth century. It was supplanted by correlation tests with more favorable statistical properties, including those developed by Pearson, Fisher, Spearman and Kendall. Unfortunately, nowhere in the literature is it explained *why* the matching method is a poor test of independence. On the contrary, there have been numerous attempts to revive the matching method in one form or another (c.f. Barton and David, 1956; Rae, 1984).

The failure of these attempts, together with the matching method's relative obscurity, may lead the reader to think superfluous a paper proving its inferiority. Surely a forgotten statistical test is a dead statistical test. Quite the opposite: Evans and his coauthors (2002), apparently completely unaware of the history of the matching method, *independently discovered it* as a means of validating an ideographic measure. Their paper has only been cited eight times (according to Google Scholar), but a second paper based on their methodology has been widely cited as establishing the validity of a different measure used in cognitive analytic therapy (Bennett and Parry, 1998).[1] One aim of the present paper is to put the final nail in the matching method's coffin.

I begin by applying Rae's theorem (1987), which proves that the Pearson correlation between *m* and rank correlation statistics like Spearman's rho and Kendall's tau goes to zero as *n* goes to infinity. I show that an inevitable consequence of Rae's theorem is the statistical deficiency of *m*. I go on to show that the deficiency of m is evident from the characteristics of its sampling distribution, which further imply that the matching method is almost always hopelessly underpowered. I also illustrate the matching method's lack of power with Monte Carlo simulation results.

A second aim of this paper is to develop statistical deficiency as a more general concept. Why should we care about terrible estimators? First, understanding deficiency may help us identify other statistical tests we ought to avoid. Second, a statistic that covaries more with sampling error than with Fisher information is so counterintuitive that understanding the phenomenon may well lead us to equally counterintuitive applications (Nelson, 2020a). Finally, I exploit the statistical deficiency of *m* to construct an indicator of the orientation of a correlation coefficient's sampling error, specifically that of Pearson's *r* computed from the same bivariate sample as *m*.

---

[1] Bennett and Parry (1998) owe their use of the matching method not to the 2002 paper but to an earlier conference presentation.

## 2     Origins of the matching method

In this section, I provide a general review of the development of the matching method. In brief, investigation of the expected number of matches among a set of *n* pairs began with one of the earliest problems in modern combinatorics, at a time when games of chance were the primary models for probability. The combinatorial principles derived in the process of obtaining a solution would be used, centuries later, to construct one of the first null hypothesis tests. Ultimately, the matching method fell into disuse, though there have been sporadic attempts over the decades to revive it in modified form.

**The hat problem.** Eighteenth-century mathematician Pierre Raymond de Montmort introduced an early combinatoric puzzle in one of the earliest comprehensive works on probability (1713). A well-known version of the rencontres problem is called the *hat problem*: If *n* gentlemen leave their hats at hatcheck as they arrive at a party, and each receives a hat at random as he departs, what is the probability that no man will depart with his own hat?[2] Montmort (and later Nicholas Bernoulli) solved the problem, showing that when *n* exceeds about 4, the probability approaches a limit equal to the inverse of *e*, the root of the natural log.

Of course, the hat problem is not really about hats but about any set of elements, and the number of those elements that stay in the same position after randomly rearranging the set. In other words, it is about comparing a pair of *random permutations* of a set of distinct elements, in terms of the number of *fixed points* between the permutations. When there are no fixed points—every element moves out of its original position after randomization—this is called a *complete derangement*. Permutations with fixed points numbering between 1 and n – 1 are called *partial derangements*.

---

[2] Another version: If a deck of cards is shuffled until random, what is the probability that the shuffle moves every single card to a new position (order) in the deck?

The original matching problem considers only the probability of obtaining a complete derangement after randomization. Later, mathematicians became interested in a generalized version of the problem[3], one that asks, "What is the probability that the number of random correct matches *m* between guests and hats is at least *k*?" Leonard Euler (eponym of *e*) followed with a general solution for determining the probabilities of exactly *m* men receiving the correct hat, for any *m*.[4] He proved (as described in Evans, et al, 2002) that the answer for any *m* converges to the probability of randomly drawing *m* from the Poisson distribution with parameter $\lambda = 1$. As this elegant solution implies, the number of men leaving with their own hats is a random variable distributed approximately Poisson(1). Thus, the solution to the original matching problem, the probability of obtaining a complete derangement at random, is the inverse of *e* because the probability of drawing a zero value from this distribution at random is $e^{-1}$. Significantly, Euler's solution applies (approximately) regardless of the number of guests.

I qualify the preceding statement with "approximate" because the precise answer actually does depend on the total number of partygoers (or, more generally, the size of the set) *n*, in that the distribution of *m* only converges to Poisson($\lambda = 1$) as *n* increases. However, the rate of convergence is quite high, as we can see in Table 1.

---

[3] Known as the rencontres ("encounter") problem, or the problem of coincidences

[4] He proved it using a version of the inclusion-exclusion principle: We want to know the number of permutations of hats and guests in which no guest is matched with his own hat. We can start by taking the number of all permutations, *n*!, minus the single permutation where all *n* hats match, minus the number of permutations where at least *n* – 1 hats match, minus the number where at least *n* – 2 hats match, and so on through the number permutations where at only 1 man leaves with his own hat. But in making these exclusions, we have subtracted too many times: when we subtracted the number of permutations with at least *n* – 1 matching hats, for example, that number included the permutation with all hats matching, which we had already subtracted. To get back to the correct number, we must now add back in the permutations subtracted too many times. The formula we end up with is a sequence of additions and subtractions, hence the "inclusion-exclusion" principle.

Table 1

|   | P(m = k, n = n) | | | | Poisson(1) |
|---|---|---|---|---|---|
| k | n = 4 | n = 5 | n = 6 | n = 7 | n = ∞ |
| 0 | .3750 | .3667 | .3681 | .3679 | .3679 |
| 1 | .3333 | .3750 | .3667 | .3681 | .3679 |
| 2 | .2500 | .1667 | .1875 | .1833 | .1839 |
| 3 | .0000 | .0833 | .0556 | .0625 | .0613 |
| 4 | .0417 | .0000 | .0208 | .0139 | .0153 |
| 5 | -- | .0083 | .0000 | .0042 | .0031 |
| 6 | -- | -- | .0014 | .0000 | .0005 |
| 7 | -- | -- | -- | .0002 | .0001 |

The probabilities in Table 1 are computed as (Evans et al, 2002, p. 387):

$$P(m = k, n = n) = \frac{1}{k!} \cdot \left(1 - \frac{1}{1!} + \frac{1}{2!} - \frac{1}{3!} + \cdots + \frac{(-1)^{(n-k)}}{(n-k)!}\right)$$

where *n* is the number of set elements, *m* is the number of fixed points and $k \leq n$. Note that departures from the Poisson distribution (far-right column) are quite small even for relatively small *n*. To obtain cumulative probabilities, simply take the summation of this quantity for all *k* through *n*, inclusive.

**The matching method.** We can bring these combinatorial principles into the realm of statistics by recognizing some parallels between the two domains. First, counting fixed points (matches) between a pair of permutations is a way of quantifying the permutations' similarity, just as when we compute the concordance of nominal data with Kendall's *W* or the correlation of ordinal data with Spearman's *rho*. Second, a pair of permutations of the integers from 1 to *n* is equivalent to a sample of paired, ranked data (without ties). Third, while randomness is not the same as independence, a consequence of filling a pair of vectors with randomly paired observations from two independent populations is that the vectors will be uncorrelated in the long run.

Taken together, these parallels between combinatorics and statistics imply that the number of fixed points between paired observations *X* and *Y* on nominal or ordinal variables, or on continuous variables converted to ranks, may be interpreted as a random variable with distribution approaching Poisson($\lambda = 1$) when *X* and *Y* are mutually independent. This is the rationale by which early-twentieth-century methodologists constructed a null hypothesis significance test (NHST) for concordance called the matching method. Vernon (1936b) credited[5] its initial development to European psychologists, graphologists, and parapsychologists[6], including Alfred Binet, Rudolf Arnheim, Otto Bobertag and Werner Wolff.

As an example of how the NHST works, suppose a researcher wants to test the theory that pet owners tend to resemble their pets. Her test might consist of her[7] blindly attempting to match a set of pet owners' photos with a set of photos of their respective pets. In this case, she would compute test statistic *m* as the number of correct matches. A simple NHST of whether her matches predicted pet-owner associations better than chance may be conducted by computing the probability of randomly drawing the observed value of *m* from a Poisson($\lambda = 1$) population (as in Table 1). If that *p*-value is less than the Type I error rate, then the observed value of *m* is found to be statistically significant and the researcher rejects the null hypothesis that, in this case, the appearances of people and pets are unrelated.

A version of the matching method was later developed for ordinal data (Rae, 1984) as an alternative to rank correlation statistics like Spearman's *rho*. Extending the example above, the same researcher might reframe her experimental hypothesis to posit that more attractive people have more attractive pets. She could then test the hypothesis by ranking owners' photos in terms of perceived attractiveness, and separately (blindly) ranking pets' photos by the same criterion. Now *m* is the number of owners who received the same attractiveness-rank as their dogs. A significant value of *m* is interpreted as rejecting the null that the distribution of relative pet attractiveness is random with respect to owners' relative attractiveness.

---

[5] However, the earliest use of the method as a hypothesis test was probably in 1819, when Thomas Young applied it to assess the relatedness of two languages (Basque and Egyptian) by matching words between them.

[6] Three fields that, in the 1930's, were closely related due to hypothesized links among mental, physical and behavioral traits. A disturbing thread of this work used the matching method to investigate whether people can "accurately" stereotype others. Participants would attempt to match strangers' images or handwriting to traits including social class, trustworthiness and in one case (Roback, 1935) "Jewishness."

[7] In this case, because the design includes only the researcher herself matching photos, the experimental hypothesis is not technically whether the pets and owners look alike but whether the researcher has the ability to detect physical similarities between owners and pets, which assumes that there is a physical similarity to detect.

The matching method sets up an NHST quite different from the classical Neyman-Pearson construction. If we want to determine whether the parameter of interest (here, the population correlation coefficient between two variables) is not zero, Neyman and Pearson advise us to estimate the parameter from a bivariate random sample, to make certain assumptions about its sampling distribution (in the parametric case), and finally, to compute the probability that the observed value of the statistic would be generated at random from the assumed sampling distribution with expected value zero. The matching method, in contrast, requires estimating a parameter ($\lambda$) separate from the parameter of interest ($\rho$), and from that estimate inferring the probability that the process generating the bivariate sample was truly random. The non-randomness of one set of independent and randomly sampled observations with respect to another implies a non-zero correlation between variables, so we can infer that this is also the probability that the population correlation is zero.

Unlike the conventional NHST, where the null value need not be zero, the matching method is only used to test the so-called nil hypothesis that any observed relationship between the two variables is due entirely to random error. There is also little need to be concerned with sample size for statistical power: because a Poisson variable with $\lambda = 1$ is distributed the same at any sample size, power is at its maximum so long as the sample is large enough to approach Poisson(1) when the null is true. The key point here is that the classical NHST and the matching method NHST are based on estimators of two different parameters and test two different null hypotheses, although both methods can be used to assess the probability that $\rho = 0$ (Barton and David, 1956).[8]

This is a very clever way to assess independence, particularly in an era before significance testing methods were well-established. The matching method at one time featured in a fairly large literature, both theoretical and applied. Once alternatives were found, new papers grew sporadic.[9] Most work in the last several decades has sought alternative applications or modified formulations,[10] but the original matching method has seemingly fallen into disuse, if not total ignorance, among applied social scientists.[11]

---

[8] Barton and David (1956) refer to "a distortion of the random sequence" (p. 69) as the subject of the hypothesis being tested by the matching statistic.

## 3  Rae's theorem and the matching method

In 1987, Gordon Rae proved a highly consequential theorem for the matching method. Rae sought to define the relationships between *m* and other indices of bivariate rank correlation. He proved, using Hubert's procedure (1979) and "tedious but straightforward algebra" (p. 376), the following relationship between *m* and another rank correlation, in the special case of independent populations:

$$Corr(m, r_s) = \frac{1}{\sqrt{n-1}}$$

where *Corr*() indicates the Pearson correlation and $r_s$ is the Spearman rank correlation. Rae also showed that the Pearson correlation between Kendall's $\tau$ and *m* is a function of the same relationship. In a later paper with John E. Spencer (1991), Rae would show that the same relationship exists for indices of rank correlations among more than two sets of ranks under the true nil hypothesis:

$$Corr(M, \bar{r}_s) = Corr(M, W) = \frac{1}{\sqrt{n-1}}$$

---

[9] Though, over the course of a century, sporadic publications add up to a substantial literature, one this abbreviated review cannot possibly do justice. In addition, I do not here even scratch the surface of the vast literature on combinatorial matching unrelated to psychometrics.

[10] Gillet (1985, 2001), for example, has developed a matching test procedure that classifies matches not as correct or incorrect, but as supporting the experimental hypothesis or contradicting it. The test, which does not involve the Poisson distribution, resembles the Fisher exact test.

[11] The method is so obscure, in fact, that it was apparently independently rediscovered in nearly a century later by Evans et al (2002), as mentioned in the introduction.

where $M$ is the matching statistic computed for $k$ sets of ranks ($k \geq 2$), $\overline{r_s}$ is the mean of the $k(k-1)/2$ values of Spearman's *rho* computed between each pair of ranks, and $W$ is Kendall's coefficient of concordance for the same.[12]

Rae described the first relation as "rather surprising" (1987, p. 377) because the correlations between $m$ and $r_s$ and between $m$ and $\tau$ must go to zero as $n$ goes to infinity, even as the correlation between $r_s$ and $\tau$ goes to 1. He remarked (p. 377), "Thus, from the standpoint of asymptotic relative efficiency, the index [$m$], based on the matching paradigm, does not possess the same properties as Spearman's or Kendall's measures." Rae and Spencer (1991) similarly described the second relation as "interesting" (p. 162). In neither paper was their curiosity pursued further.

What I find both surprising and interesting is that Rae's correlation formula is identical to the formula (David and Mallows, 1961) for the standard error of the sample correlation coefficient with distribution "determined by the $n!$ pairings [i.e., permutations] of two given sets of numbers...when all such pairings are equally probable [i.e., $\rho = 0$]" (Pitman, 1937), also known as the null hypothesis for Spearman's *rho*. Accordingly, I observe that both the correlation between $m$ and $r_s$, and the standard error in $r_s$, shrink at the identical rate and to identical levels relative to increasing $n$. Contrast this with the simultaneous increase in the correlation between $r_s$ and $\tau$, which goes to 1 with infinite $n$, a consequence of the infinite increase in Fisher information (relative to error) in each. It is also significant that the conditional relationship between $r_s$ and $m$ is entirely linear (Rae and Spencer, 1991), so there can be no hidden, nonlinear covariability between $r_s$ and $m$ not captured by Pearson's $r$.

The relevant implications of Rae's theorem are these: that $m$ shares a substantial component of random error variance with $r_s$ that diminishes only gradually with increasing $n$—obvious from the variance of $r_s$ and its covariance with $m$ being identical; and that the Fisher information contained in $m$ is negligible at any sample size and rapidly approaches an absolute limit—this is clear because, otherwise, its correlation with $r_s$ could never be 0, *especially* when $n$ is infinite and $r_s$ contains the complete Fisher information about $\rho$. Together, these implications imply one further: that the sampling distribution of $m$ is highly related to the error variance of the sampling distribution of $r_s$ but only negligibly related to the expected value, which is a function of Fisher information.

---

[12] Although Rae had already established in his previous paper (1987) that this must be true generally, albeit without providing the detailed proof in the second paper (1991).

Note the distinction between sampling error variance, sampling error, and standard error. *Sampling error variance* describes the expected deviations of an estimator from its parameter, from one random sample to the next, across infinite samples. It arises because each sample represents observations on a different subset of the entire population. Sampling error variance is a property of an estimator's sampling distribution. *Sampling error* is a property of an individual estimate computed from a particular sample: it is the size and direction of the estimate's deviation from its parameter. It is the short-run result of the long-run error variance.

When samples consist of observations drawn at random from a population, sampling error is random. In contrast, standard error is the (square root of the) total variability in a sampling distribution, a combination of both sampling error and non-sampling error (bias from non-random factors like measurement error, sampling from multiple populations due to misclassification of subjects or misspecification of constructs, human error in data entry, etc.). In other words, Rae's theorem describes how $m$ covaries with rank correlation coefficients in the absence of, or controlling for, all non-random sources of error—though only in the special case of a true nil hypothesis.

Figure 1A uses Monte Carlo simulation results to illustrate what Rae's theorem predicts, that the correlation between $m$ and $r_s$ goes to 0 with increasing sample size. The joint distribution of $r_s$ with $m$ is in red, computed from 100,000 simulated samples, of size $n = 10$, drawn at random from a null population. The equivalent joint distribution with samples of size $n = 50$ is in blue. Both joint distributions have been jittered for clarity.

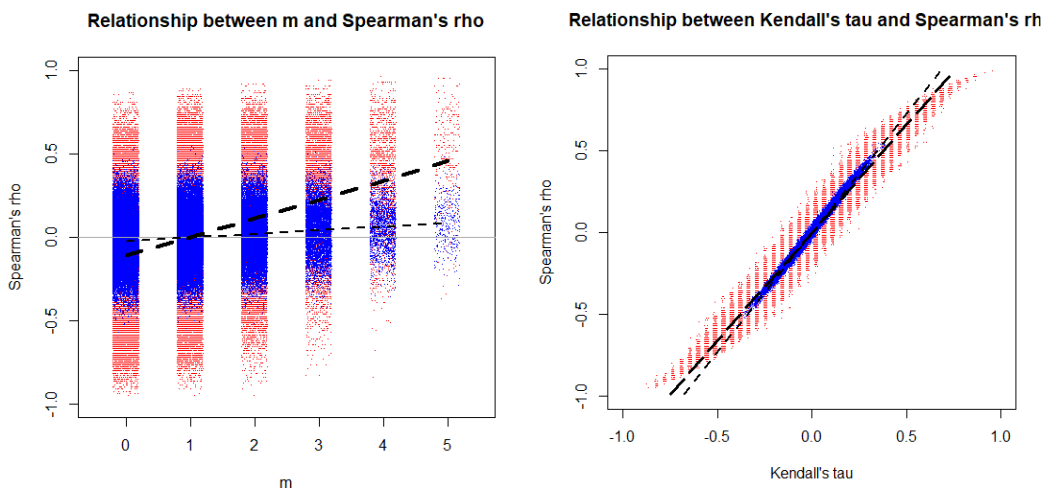

Figure 1. The relationship between *m* and Spearman's rho (left) under a true null is almost entirely due to their shared sampling error variance. We can see this in terms of the slope of the regression line when sample size is $n = 10$ (long dashes) and sampling error is large (red dots) compared to the slope when sample size is $n = 50$ (short dashes) and sampling error is small (blue dots). Less sampling error equates with less association between the two statistics. Contrast this with the relationship between Spearman's rho and Kendall's tau (right), which changes little as a function of sampling error, indicating their shared information about the population correlation is considerable.

At the smaller sample size, where sampling error variance is high for both statistics, the two statistics are moderately related. At the larger sample size, where the error in Spearman's rho is far less, they covary far less: the slope in the regression line when $n = 50$ (short dashes, $\beta = .15$, $r^2 = .02$) is less than half the slope when $n = 10$ (long dashes, $\beta = .34$, $r^2 = .11$). This is consistent with my reinterpretation of Rae's theorem: because *m* shares far more error variance than information with Spearman's rho, increasing information and decreasing error in the latter can only reduce their association.

Rae's theorem also implies that increasing sample size will have the opposite effect on the joint distribution of Spearman's rho with Kendall's tau, since these two estimators are related mostly through their respective information about $\rho$. Figure 1B confirms the prediction: when sample size goes from $n = 10$ (red) to $n = 50$ (blue), the regression slope ($n = 10$, $\beta = .98$, $r^2 = .97$; $n = 50$, $\beta >$

.99, $r^2$ = .99) barely changes. To the extent the relationship does change, it grows stronger with more information.

Figure 1A also suggests the statistical power of the matching method may be a concern when the null is not true. Note that the width of the sampling distribution of *m* spanning the *x*-axis is essentially the same at either sample size. This is a further implication of Rae's theorem: a limit on information is a limit on precision. Meanwhile, the sampling distribution of Spearman's rho narrows considerably along the y-axis narrows considerably. If this contrast holds under a false null, the matching method will have very low power indeed.

Unfortunately, Rae's theorem applies only when $\rho = 0$, so describing the statistical power of the matching method requires taking a different tack. In the next section, I answer the question of what it is about *m* that makes it so insensitive to increasing Fisher information in a sample. The distinction between Rae's approach and my own is that, where his work (inadvertently) established the limited information of *m* as an emergent property of its joint distribution with rank correlation coefficients, I deduce it as an inherent property of the unconditional sampling distribution of *m*. From this more general departure point, I am able to reestablish everything implied by Rae's theorem about the null case and more.

## 4     The statistical deficiency of *m*

The extremely low statistical power of the matching method follows directly from the fact that *m* is asymptotically distributed Pois(1), a limit that it rapidly approaches in very small samples (Barton, 1958; Evans et al, 2002). To see how the one fact implies the other, consider that the parametric (Fisher) information of a statistic is an inverse function of its standard error, the standard deviation of its sampling distribution. This is why we generally prefer to compute parameter estimates from larger samples: we increase sample size in order to decrease the sample estimator's standard error and increase its Fisher information, which in turn increases our power.

Surprisingly, this is not the case for *m*: increasing sample size only pushes *m*'s sampling distribution closer to the limiting form Pois(1), which, of course, has a fixed standard deviation of 1. Therefore, as we increase sample size beyond a few observations, we only move the standard error of *m* closer to 1. An absolute limit on standard error means an absolute limit on Fisher information and therefore on power. Figure 2 illustrates this idiosyncrasy of *m* by contrasting its sampling distribution with that of Spearman's rho when n = 10 and 50. While the distribution of rho narrows considerably around its expected value of 0 at the larger sample size, the distribution of the former remains unchanged.

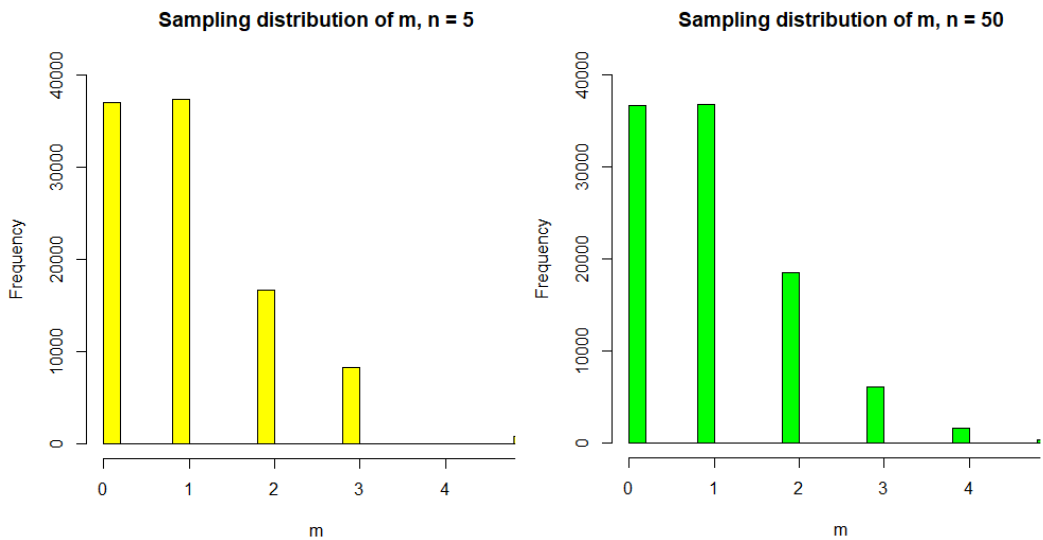

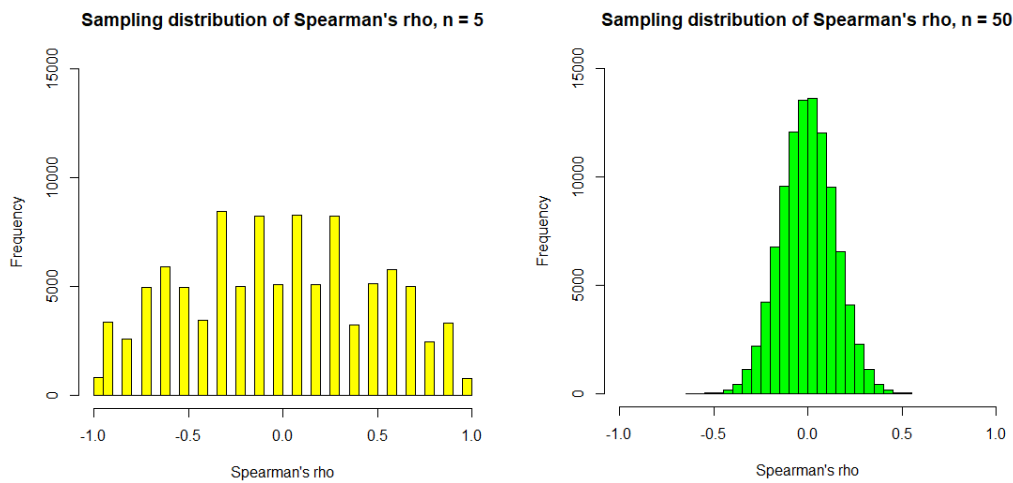

Figure 2. Increasing sample size from n = 5 (yellow) to n = 50 (green) results in an extreme narrowing of the sampling distribution of Spearman's rho (bottom) but virtually no change in the sampling distribution of *m* (top).

The distributions in Figure 2 are for $\rho = 0$ and so merely reiterate what was deduced from Rae's theorem, and illustrated with different Monte Carlo simulations, in the last section. In contrast to Figure 1, however, the distributions of *m* in Figure 2 (top) are independent of the distributions of Spearman's rho (bottom). Having abandoned the constraints of Rae's theorem, I can now describe the limits on the power of *m* when the null is false.

Consider that *m* is the sum of *n* Bernoulli variables: it can be computed by coding each pair of ranked observations as a 1 when the ranks match and a 0 when they do not, and then taking the sum. It is well known that the sum of *n* Bernoulli variables is distributed approximately Poisson with $\lambda = \Sigma p_i$, where $\Sigma p_i$ is the sum of the probability of success across each of the *n* individual variables.[13] We can deduce the following: when $\rho = 0$, *m* is approximately Poisson with $\lambda = 1$.

---

[13] This assumes the population correlation is not large. However, large correlations are the exception, not the rule, in social science research. When we do encounter large correlations, we rarely have difficulty distinguishing them from 0 by definition.

When $\rho > 0$, $m$ is approximately Poisson with $\lambda > 1$, because the expected number of matches (and therefore $\Sigma p_i$) only increases with the population correlation. Since the standard deviation of a Poisson-distributed variable is equal to the square root of $\lambda$, the standard error of $m$ must also be greater than 1 when $\rho > 0$. Therefore, at any but the smallest sample sizes and for any positive population correlation value, $m$ will always have limited power to reject a false null.[14]

We can also see this empirically through Monte Carlo simulations. I conducted a two-factor simulation, with 5 levels of sample size ($n$ = 10, 30, 50, 100, 200) and 9 levels of population correlation ($\rho$ = -.7, -.525, -.35, -.175, 0, .175, .35, .525, .7), for a total of 45 cells. Per cell, I simulated randomly drawing 100,000 bivariate standard normal samples and computed the proportion of samples for which the null hypothesis would be rejected by the matching method, asymptotic $\alpha$ = .0417. Figure 3 reveals that power is positively correlated with $\rho$, such that power is lowest when $\rho$ = -.7 and rises until $\rho$ = .7. We can attribute the lower power for negative values of $\rho$ to $m$ being a count variable: $m$ can never be less than zero, so it gives comparatively little information about negative covariance. (See also Footnote 13.)

---

[14] When $\rho < 0$ the standard error of $m$ is less than 1, but it will always be large relative to the standard error of $r$. For example, suppose the population correlation is extremely negative so matches are highly unlikely, say $p$ = .06 per paired observation. Then, for $n$ = 5, $\Sigma p_i$ = .3 and the standard error of $m$ is .548. The standard error of the sample correlation is already .5 when $\rho$ = 0 and so will be far less at the extremely negative value we have supposed. See also the negative correlation region in Figure 3.

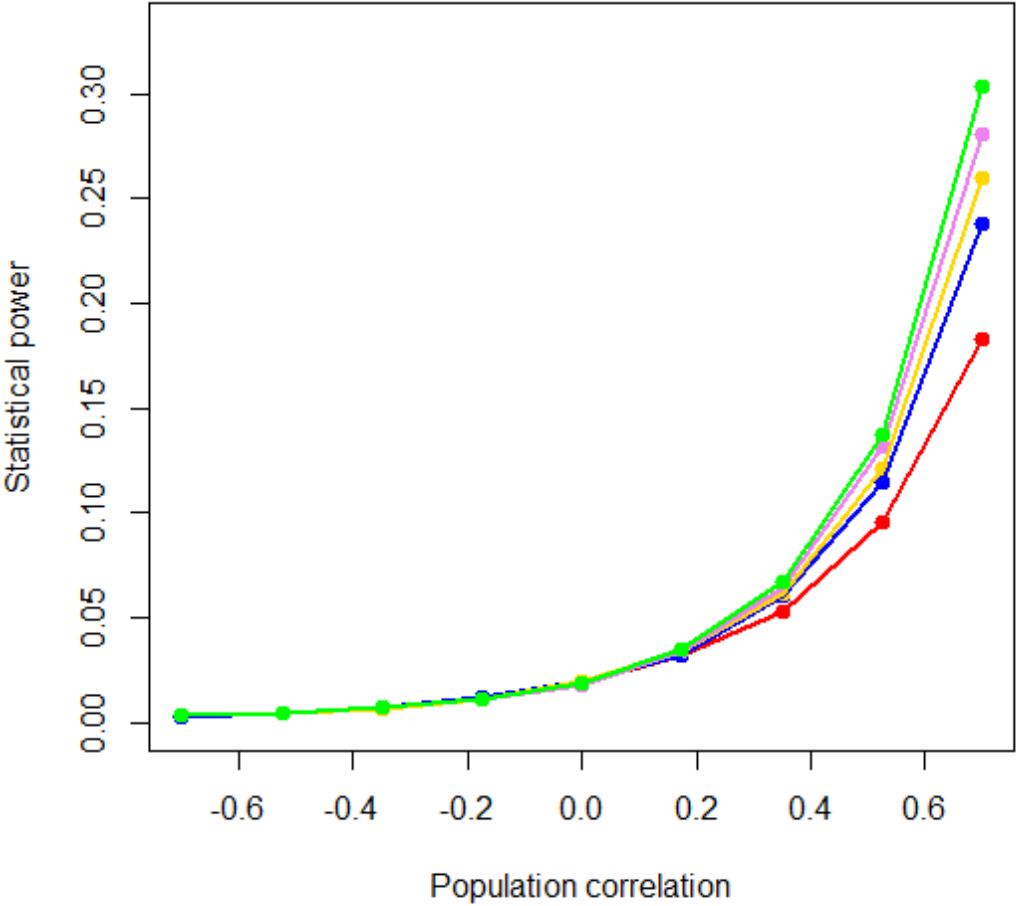

Figure 3. Across all sample sizes (red = 10, blue = 30, orange = 50, violet = 100, green = 200) in the Monte Carlo simulations, statistical power is negligible when $\rho < 0$. As $\rho$ increases, so does power, but only very gradually until population correlations are large, when rate of increase diverges by $n$.

Under a true null, power is essentially static across sample sizes, just as in Figure 2. When $\rho > 0$ and held constant, power increases with $n$ but only slowly and only up to a limit. For $\rho = -.7$, for example, the increase in power from $n = 10$ (red) to $n = 30$ (blue) is about the same as the increase from $n = 30$ to $n = 200$ (green)—and even at such a large sample size, power is still only about 30 percent. When sample size is held constant, power increases with $\rho$ very slowly until $\rho$ becomes large. For large $\rho$, the distribution of $m$ is no longer well approximated by Poisson (see Footnote 12).

What is it about m that makes it so insensitive to information? One thing that distinguishes $m$, as we have seen, is that it is highly *insufficient*. A sufficient statistic (Fisher, 1922) coveys all of the Fisher information in a sample about its parameter, whereas $m$ has a definite limit on the amount of Fisher information it can take from any sample. We can also say that $m$ is highly *inconsistent*. A consistent statistic is one that inevitably converges to the true parameter value as $n$ goes to infinity. This means that a consistent statistic's sampling distribution ultimately narrows to the parameter value, but we have seen that $m$'s sampling distribution can never be narrower than a single standard deviation, except in certain special cases. Finally, $m$ must be highly *inefficient*. The efficiency of a statistic is the rate at which it asymptotically converges to the true parameter value with increasing $n$. We know that $m$ will *never* converge to the parameter value at *any n*.

Sufficiency, consistency, and efficiency are the main criteria by which the performance of an estimator is judged. The estimator $m$ is not just low on each metric, it is characteristically stunted. The absolute floor on sampling error is particularly surprising as it appears to contradict a common interpretation of sampling theory, as offered by Biemer and Lyberg (2003): "...as we increase the sample size...sampling error becomes smaller and smaller. ...Thus, sampling error can be made as small as we may wish (or can afford) to make it by manipulating the sample size" (p. 37). I call a non-arbitrary[15] estimator with these attributes a

---

[15] An arbitrarily deficient estimator would be computed through a procedure that, as a rule, intentionally uses partial or no Fisher information from the sample. Examples include an estimator with constant value or one computed on an arbitrary subset of the sample observations (e.g., choosing a single observation at random from the sample to serve as the estimate). It is also trivial to find pathological distributions, like the Cauchy distribution, for which non-deficient estimators like the sample mean display deficient behavior.

*deficient* statistic, and *m* appears uniquely deficient among statistics actually applied in the literature.[16]

## 5    Deficient statistics as sampling error indicators

The relevant consequences of statistical deficiency may not be immediately apparent, so I offer the following thought experiment. Suppose we compute both *m* and Spearman's *rho* ($r_s$) for a single, bivariate normal (ranked) sample of $n = 10$, assuming zero measurement error. There is very little Fisher information in such a small sample, especially as a proportion of random error variance. We would of course like to interpret the distance and direction of *r* from zero as an indicator of the magnitude and sign of the parameter $\rho$, but we know that its value is substantially a result of sampling error. For the same reason, the distance of *m* from 1 (the expected value of *m* for independent populations) cannot be interpreted as reliably indicating anything other than the magnitude of its sampling error.

    Now suppose we repeat these calculations for a bivariate normal sample of $n = 50$. There is now a great deal of Fisher information in the sampling distribution of $r_s$ about $\rho$. We may now interpret the value of $r_s$ as indicating both the magnitude and the sign of the parameter. In contrast, the sampling distribution of *m* still contains little more Fisher information about $\rho$ than it did when $n = 5$. Any deviation in *m* from 1 still mostly indicates the extent that *m* has been perturbed by random error variance. That is, whenever we observe values of *m* much greater than 1, *especially* in large samples,[17] we have little reason to believe *m* has been moved to that extreme by true variance in the parameter rather than by error variance particular to the sample.

    I noted early in this paper that a counterintuitive statistical property like statistical deficiency might lead to a counterintuitive statistical application. Such an application is suggested by Figure 1, where *m* predicts the sampling error of

---

[16] Guttman (1977) notes to his readers after making a similarly sweeping claim that, of course, he could not possibly review the entirety of the literature. He encouraged readers to submit counterexamples, as do I.

[17] To further illustrate this point, Monte Carlo simulations (see Appendix) show that when conventional tests have 80% power, the matching method has power of approximately 26% at *n* =10, 10% at *n* = 30, and only 5% at *n* = 100.

Spearman's rho under a true null. The Monte Carlo simulation results illustrate a simple syllogism: if the value of *m* mainly indicates its sampling error, and if *m* mainly covaries with $r_s$ through sampling error, then the value of *m* is an indicator of a component of the sampling error in $r_s$. In other words, the counterintuitive application of *m* is that we can make inferences from *m* about the probability of whether $r_s$ is an overestimate or an underestimate of $\rho$.

At least, Figure 1 implies this is true when $\rho = 0$. The properties of *m* under a false null, discussed above, imply that it may also be possible to make limited inferences about sampling error in $r_s$ when $\rho$. To test this, I conducted another series of Monte Carlo simulations. For each of the same population correlation values as in Figure 3 ($\rho$ = -.7, -.525, -.35, -.175, 0, .175, .35, .525, .7), I simulated drawing 100,000 samples of size *n* = 15 from a bivariate standard normal distribution. I converted observed values to ranks and computed *m* for each sample. I then used the location of *m* relative to its expected value under a true null (less than, equal to, or greater than 1) to predict whether the sample correlation was an overestimate of the population correlation.

Figure 4 displays the results, with one important deviation from earlier plots: the correlation coefficient predicted by *m* is Pearson's *r*, not Spearman's rho. I made this change primarily because r is used far more often in research than is $r_s$, or any other rank correlation coefficient. The major motivation for using Spearman's rho earlier was to extend Rae's theorem. Now that the deficiency of m has been established separate from Rae's theorem, the analysis is no longer limited to rank correlations, just as we are no longer limited to cases of a true null. Transforming to ranks need only be a step in computing the sampling error indicator *m* for *r*.

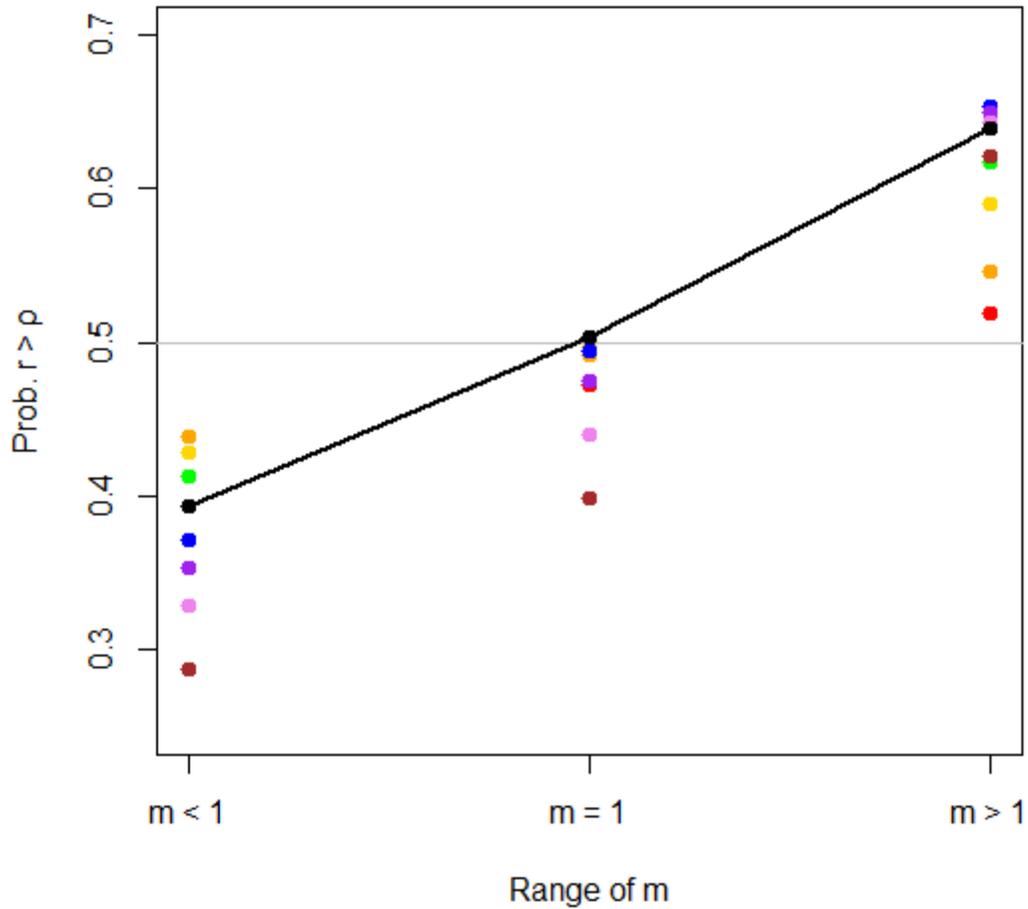

Figure 4. Whether the value of m computed from a sample of size $n = 50$ is below or above its expected value under a true null, $E(m \mid \rho = 0) = 0$, predicts whether Pearson's $r$ computed from the same sample is an underestimate or an overestimate of $\rho$. The degree of discrepancy is related roughly to the distance of $\rho$ from nil (red = -.7, orange = -.525, gold = -.35, green = -.175, black = 0, blue = .175, purple = violet = .35, brown = .525, gray = .7). The black diagonal line merely traces the probabilities for $\rho = 0$ across the three ranges.

Based on Figures 4 and 5, we can conclude that observing $m < 1$ implies a better than even chance that the value of $r$ computed from the same sample is an underestimate, and we can conclude the opposite when $m > 1$, when $n = 15$ and so long as we can assume $-.7 \leq \rho \leq .7$. When $m = 1$, the probability that $r > \rho$ peaks at .5 when $\rho = 0$ (black points and line), so we can conclude that $r$ is not more likely to be an overestimate than an overestimate. Figure 5 shows the same for $n = 50$. In general, to make the same inference with larger sample sizes requires that we assume a population correlation closer to $\rho = 0$, while smaller samples permit a wider range of possible parameter values.

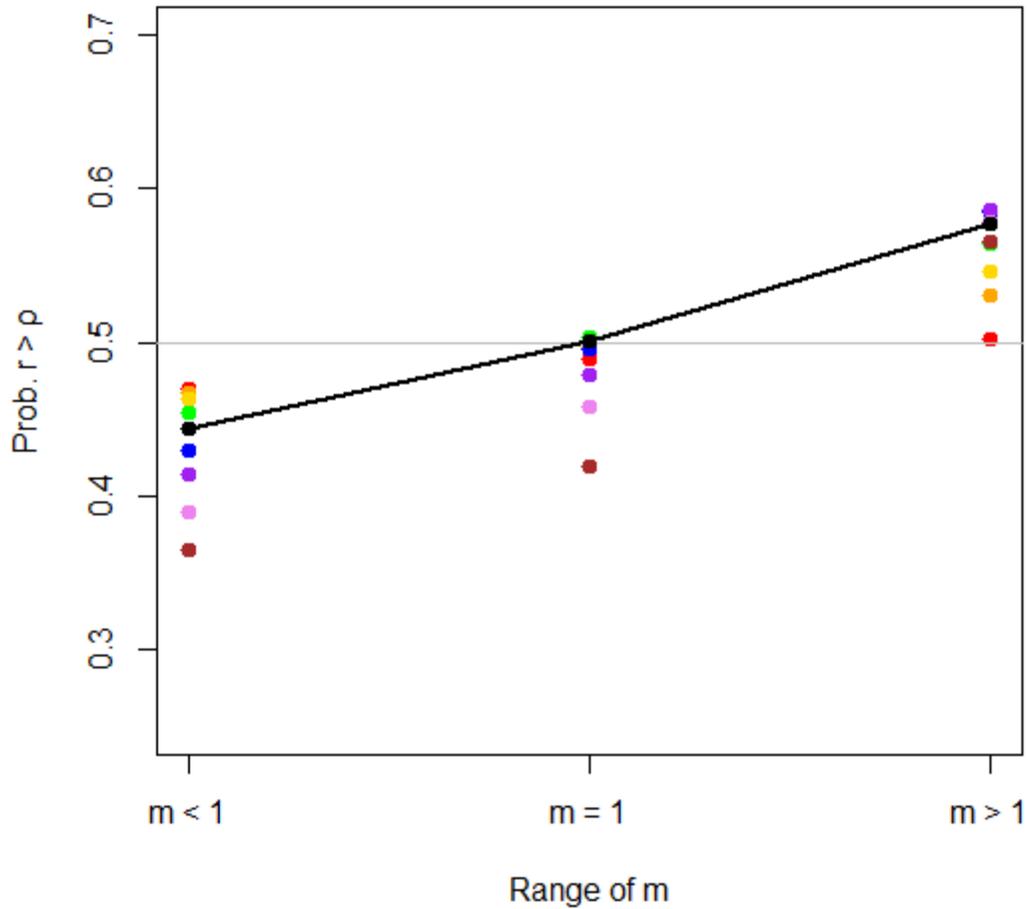

Figure 5. Whether the value of m computed from a sample of size $n = 50$ is below or above its expected value under a true null, $E(m \mid \rho = 0) = 0$, predicts whether Pearson's $r$ computed from the same sample is an underestimate or an overestimate of $\rho$. The degree of discrepancy is related roughly to the distance of $\rho$ from nil (red = -.7, orange = -.525, gold = -.35, green = -.175, black = 0, blue = .175, purple = violet = .35, brown = .525, gray = .7). The black diagonal line merely traces the probabilities for $\rho = 0$ across the three ranges.

# 6    Conclusion

This paper explicitly demonstrates three main points: the matching method is underpowered; *m* is a deficient statistic for the population correlation; and *m* can be used as an indicator of the direction of sampling error in sample correlation coefficients including Pearson's *r*, given certain assumptions. The first point is purely methodological—avoid using the matching method or statistical tests with similar limitations. The second is purely theoretical—*m* represents a class of estimators with properties, uses, and misuses previously undocumented and unanticipated. The third contributes both to methods and theory.

In principle, an estimator of sampling error's sign could be used to construct confidence intervals that are vectors, with terms indicating both the probable distance and the probable direction of the parameter from the point estimate. For example, the expression $\{.2 \leq \rho \leq .25)$ might signify that the population correlation lies between .2 and .25, with some specified probability, and will be closer to .2 than to .25 more than half of the time. However, the proposed statistic is best considered a crude prototype at this point. Before we can use it and interpret it to make claims with statistical validity, its capacities and limitations require a foundation more solid than I have provided in this paper.

More interesting to me is a fourth, implicit point: we now know that it is possible to make inferences about the sampling error in a statistic using only the observations from which the statistic has been computed. This result is consistent with a framework I have proposed elsewhere for a theory of sampling error estimation (Nelson, 2020a), although the method introduced here only estimates the sign of the sampling error. In a forthcoming paper (Nelson, 2020b), I advance that framework toward a real theory and additional applications.

# Appendix

Figure A displays the results of a simulation with 4 X 4 levels. For each of four sample sizes ($n$ = 10, 30, 50, and 100), I simulated drawing 100,000 random samples from bivariate normal populations with correlations corresponding to four levels of power (50%, 60%, 70%, and 80%) for a conventional NHST.[18] I converted sample observations to ranks so that $m$ could be computed. Finally, I calculated the proportion of samples (at each level of $n$ and power) for which $m > 3$, corresponding to an asymptotic $p$-value of $.04 < p < .05$ for the matching method NHST.[19]

---

[18] Exact correlations for each sample size were: $n$ = 10: $r$ = .62, .67, .72, .78; $n$ =30: $r$ = .36, .40, 0.442, .49; $n$ = 50: $r$ = .277, .311, .346, .386; $n$ = 100, $r$ = .196, .221, .247, .277.

[19] It should be noted that the nominal power levels are given for rejecting a continuous null hypothesis—e.g. 50% power means there is a 50% probability that the Pearson's $r$ NHST would reject the null hypothesis that the population correlation between a pair of continuous variables equals zero. This is because the simulation mimics the case of an underlying continuous joint distribution. The nominal power levels for rejecting a discrete null hypothesis, e.g. for the Spearman's rho NHST, would be slightly lower than those given.

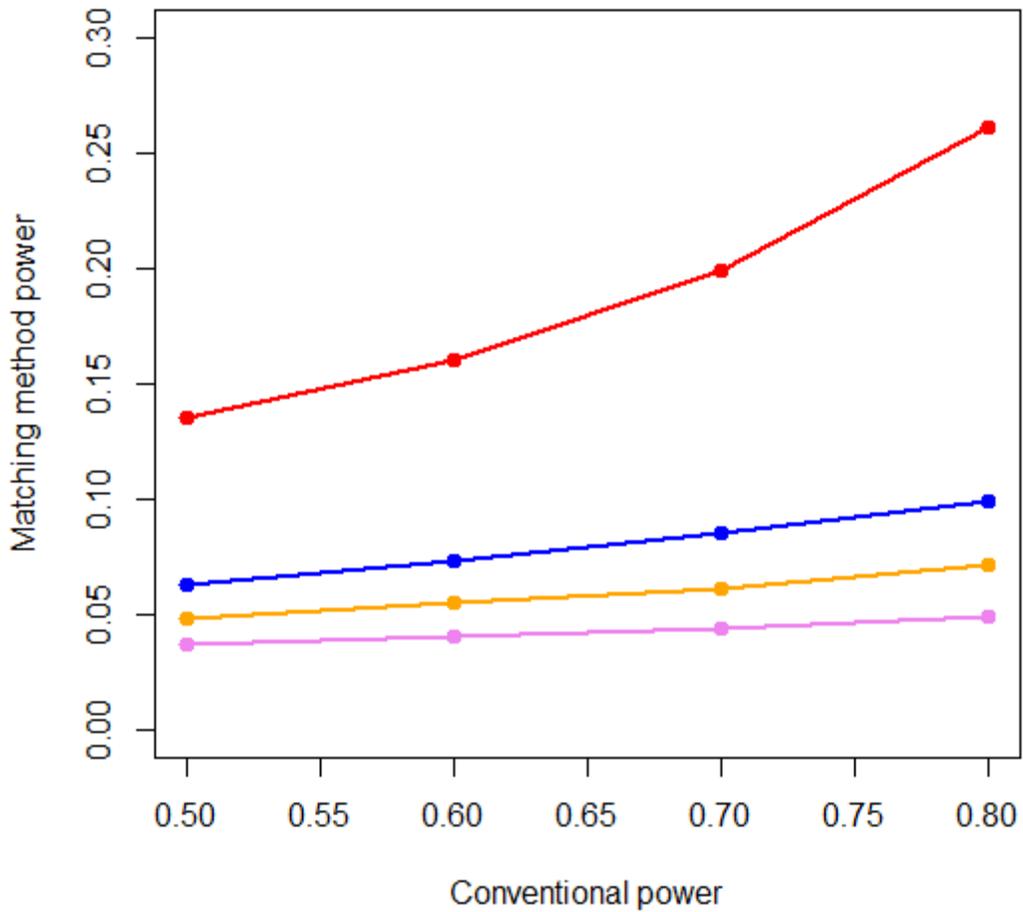

Figure A. Each line represents the probability of rejecting a false null by the conventional NHST (*x*-axis) and the matching method NHST (*y*-axis) with a certain sample size: $n = 10$ (red), 30 (blue), 50 (orange), and 100 (violet) and population correlation (see Footnote14). The graph may be read as, for example, "When $n = 10$ and the conventional NHST has 50% power to reject the null, the matching method has less than 15% power." The graph overall shows that, if the conventional test has at least 50% power, then the matching method will have far less power whatever the sample size, and by a larger margin for large *n*.

This is a genuinely counterintuitive result: the matching method's relative power only decreases as sample size increases from n = 10 (red) to n = 100 (violet).